\documentclass[conference]{IEEEtran}
\usepackage[noadjust]{cite}
\usepackage{amsmath,amsthm,graphicx,cite,booktabs,amssymb,gensymb, color, xcolor, fancyhdr,float,perpage,tikz}
\usepackage[linesnumbered,vlined,ruled]{algorithm2e}
\usepackage{algorithmic,array,enumerate,rotating}
\usepackage{subfigure,epsfig,dblfloatfix}
\usepackage{multirow}
\usepackage[affil-it]{authblk}
\usepackage{hyperref}
\usetikzlibrary{matrix}
\hypersetup{
	colorlinks   = true,
	linkcolor    = black,
	citecolor    = black
}

\makeatletter

\renewcommand{\@algocf@capt@plain}{above}

\newcommand{\norm}[1]{\left\lVert #1 \right\rVert}

\newcommand{\ab}{\textcolor{black}}

\makeatother
\def\bsx{{\boldsymbol{x}}}
\def\bsy{{\boldsymbol{y}}}
\def\bsz{{\boldsymbol{z}}}
\def\bsu{{\boldsymbol{u}}}
\def\bsv{{\boldsymbol{v}}}
\def\bsr{{\boldsymbol{r}}}
\def\bsH{{\boldsymbol{H}}}

\def\bpsi{{\boldsymbol{\psi}}}
\def\balpha{{\boldsymbol{\alpha}}}

\title{Poisson Image Deconvolution by a Plug-and-Play Quantum Denoising Scheme}
\author{Sayantan~Dutta$^{1,2}$,~Adrian~Basarab$^{1}$,~Bertrand Georgeot$^{2}$,~and~Denis~Kouam\'e$^{1}$}

\affil{
   {\em $^{1}$Institut de Recherche en Informatique de Toulouse, UMR CNRS 5505, Universit\'e de Toulouse, France} \\
  {\em  $^{2}$Laboratoire de Physique Th\'eorique, Universit\'e de Toulouse, CNRS, UPS, France} \\
  {\em  Emails: sayantan.dutta@irit.fr, adrian.basarab@irit.fr, georgeot@irsamc.ups-tlse.fr, denis.kouame@irit.fr}
 }

\begin{document}

\pagenumbering{gobble}

\maketitle

\begin{abstract}

This paper introduces a new Plug-and-Play (PnP) alternating direction of multipliers (ADMM) scheme based on a recently proposed denoiser using the Schroedinger equation's solutions of quantum physics. \ab{The efficiency of the proposed algorithm is evaluated for Poisson image deconvolution}, which is very common for imaging applications, such as, for example, limited photon acquisition.  
Numerical results show the superiority of the proposed scheme compared to recent state-of-the-art techniques, for both low and high signal-to-noise-ratio scenarios. \ab{This performance gain is mostly explained by the flexibility of the embedded quantum denoiser for different types of noise affecting the observations.} 

\end{abstract}


\section{Introduction}
\label{sec:intro}

In the field of inverse problems a primary and long-standing challenge is the restoration of a digital image in a wide range of applications such as denoising, deblurring, compression, compressed sensing or super-resolution. In number of applications such as limited photon acquisition, positron emission tomography, X-ray computed tomography, etc., the noise degrading the acquired data follows a Poisson model. These Poisson inverse problems have a core importance in the fields of photographic \cite{foi2005spatially}, telescopic \cite{starck2003astronomical} or medical \cite{fessler1995penalized} imaging. The estimation of the underlying hidden image from a distorted observation is often formulated as an optimization of a cost function implementing the idea of maximum a posteriori (MAP) estimator. This optimization task generally leads to a unique solution following the proximal operator \cite{bauschke2011convex} based iterative schemes \cite{beck2009fast, boyd2011distributed, yang2010fast, afonso2010fast, de2011alternating} with a suitable choice of regularizer depending on the prior statistics of the image to estimate.

The alternating direction method of multipliers (ADMM) \cite{boyd2011distributed, yang2010fast, afonso2010fast,de2011alternating} is a standard scheme which redefines the optimization problem into a constrained optimization framework. A few years back, a new procedure was introduced in this domain, enabling the use of state-of-the-art denoisers instead of the proximal operator, known as the plug-and-play (PnP) scheme \cite{venkatakrishnan2013plug}.  The PnP methods can use state-of-the-art denoisers such as dictionary learning \cite{elad2006image}, non-local mean (NLM) \cite{buades2005non}, block-matching and 3D filtering (BM3D) \cite{dabov2007image}, etc., and became popular for their very good performance in the field of inverse problems (\textit{e.g.,} \cite{azzari2016variance, rond2016poisson, azzari2017variance, sreehari2016plug, chan2016plug, ryu2019plug}). Interestingly, these PnP-ADMM methods do not require any prior information about the hidden image as a consequence of the intrinsic association between the regularizer and the external denoiser.

The state-of-the-art denoisers algorithms were in general developed for Gaussian noise, and are consequently not well-adapted to other types of noise such as Poisson. To mitigate this issue, it was proposed to approximately reformulate the Poisson model into an additive Gaussian noise by using a variance stabilizing transformation (VST) (e.g., the Anscombe transformation \cite{anscombe1948transformation, makitalo2011closed}), before employing a Gaussian denoiser. \ab{PnP schemes are also a way of converting the Poisson noise affecting the observed distorted image into another possibly Gaussian noise by decoupling the restoration and denoising steps. Methods combining VST and PnP have already been proposed in the literature.} Although these VST-based PnP schemes were shown to be very efficient for low-intensity noise \cite{azzari2016variance, rond2016poisson, azzari2017variance}, they are known to exhibit inaccuracies while dealing with high-intensity noise (\textit{i.e.,} low SNR) \cite{salmon2014poisson}. Moreover, the nonuniform behavior of the convolution operator under a VST \cite{rond2016poisson, azzari2017variance, deledalle2012compare} introduces theoretical flaws when dealing with image deconvolution.

In this work, we address this issue by embedding into a PnP-ADMM scheme a new adaptative denoiser \cite{dutta2020quantum, smith2018adaptive} designed by borrowing tools from quantum mechanics. The adaptative nature of this denoiser makes it highly efficient at selectively eliminating noise from higher intensity pixels, \ab{without relying on any statistical assumption about the noise. Its efficiency regardless of the assumption of Gaussian noise represents the main motivation of its interest in Poisson deconvolution PnP-ADMM algorithms, discarding the necessity of a VST.}

The remainder of the paper is organized as follows. A quick background review on PnP-ADMM algorithms is provided in Section \ref{sec:background}. The construction of the proposed method referred to as QAB-PnP for Poisson inverse problems is detailed in Section \ref{sec:optscheme}. Section \ref{sec:expe} presents the results of numerical experiments showing the efficiency of the proposed method. Section \ref{sec:conclusion} draws the conclusions.

\vspace{-2mm}
\section{Background}
\label{sec:background}
\vspace{-1mm}

\subsection{Alternating direction method of multipliers}
\label{sec:ADMM}
\vspace{-1mm}

ADMM is an iterative convex optimization method, primarily designed to solve a constrained optimization problem expressed as follows:
\vspace{-1mm}
\begin{equation}
\begin{array}{l}
\underset{\bsx,\bsz} {\textrm{minimize}}~~~~ f(\bsx) + g(\bsz)  \\
\textrm{subject to}~~ \boldsymbol{A} \bsx + \boldsymbol{B} \bsz = \boldsymbol{c},
\label{eq:ADMMgeneral}
\end{array}
\end{equation}
where $f$ and $g$ are two convex functions with $\bsx \in {\rm I \!R}^n$, $\bsz \in {\rm I \!R}^m$, $\boldsymbol{A} \in {\rm I \!R}^{p \times n}$, $\boldsymbol{B} \in {\rm I \!R}^{p \times m}$ and $\boldsymbol{c} \in {\rm I \!R}^p$. This problem can be solved by decoupling the associated augmented Lagrangian, defined as
$\mathcal{L}_\lambda (\bsx,\bsz,\bsu) = f(\bsx) + g(\bsz) + (\lambda/2) \norm{ \boldsymbol{A} \bsx + \boldsymbol{B} \bsz - \boldsymbol{c} + \bsu}_2^2 - (\lambda/2)\norm{ \bsu}_2^2$,
(here $\bsu \in {\rm I \!R}^p$ is the Lagrange multiplier, and $\lambda > 0$ is a penalty parameter), leading to the following steps repeated iteratively until convergence:
\begin{align}
& \bsx^{k+1} = \underset{\bsx} {\textrm{arg min}}~~ \mathcal{L}_\lambda (\bsx,\bsz^k,\bsu^k), 
 \label{eq:step1} \\
& \bsz^{k+1} = \underset{\bsz} {\textrm{arg min}}~~ \mathcal{L}_\lambda (\bsx^{k+1},\bsz,\bsu^k), 
 \label{eq:step2} \\
& \bsu^{k+1} = \bsu^k + \boldsymbol{A} \bsx^{k+1} + \boldsymbol{B} \bsz^{k+1} - \boldsymbol{c}.
\label{eq:step3}
\end{align}

\vspace{-2mm}
\subsection{Plug-and-Play framework}
\label{sec:pnp}

The idea behind PnP is to provide an elegant way of splitting the problem \eqref{eq:ADMMgeneral} into the measurement model \eqref{eq:step1} and the prior term \eqref{eq:step2} so that the latter is solved separately by a denoising method. The key benefit of this process is that the regularizer does not need to be defined explicitly because of its implicit dependence on the denoising operator. It can thus be implemented without having any specific construction of the denoiser from an optimization scheme, and has been shown to give more refined outcomes than prior based models \cite{yang2010fast}.

The PnP-ADMM scheme converges globally for any non-expansive symmetric smooth denoising operator \cite{sreehari2016plug}. Recently the fixed point convergence has been proven with weaker conditions (\textit{e.g.,} \cite{chan2016plug, ryu2019plug}), but we stress that all these algorithms were constructed for Gaussian noise.

\section{Proposed optimization scheme}
\label{sec:optscheme}

\subsection{Poisson deconvolution model}
\label{sec:poissdecon}

The objective of this work is to estimate an image from its blurred version contaminated by Poisson noise. The image formation model is governed by the Poisson process $\mathcal{P}(\cdot)$ as
\begin{equation}
\bsy = \mathcal{P}( \bsH \bsx ),
\label{eq:noisemodel}
\end{equation}
where $\bsy \in {\rm I \!R}^{n^2}$ represents the blurred and noisy observation of the desired image $\bsx \in {\rm I \!R}^{n^2}$ (without loss of generality we consider square images of size $n\times n$) and $\bsH \in {\rm I \!R}^{n^2 \times n^2}$ is a block circulant with circulant block matrix accounting for 2D convolution with circulant boundary conditions. Note that $\bsx$ and $\bsy$ images are presented as vectors in lexicographic order.

The MAP estimator provides an appealing way of estimating $\bsx$ from the observation $\bsy$ by maximizing the posterior probability:
\begin{equation}
\hat{\bsx} = \underset{\bsx} {\textrm{arg max}}~~ P(\bsx|\bsy).
\label{eq:PostProb}
\end{equation}

The maximization task \eqref{eq:PostProb} can be reformulated equivalently by considering $- log(\cdot)$ element wise and implementing the Bayes' theorem as
\begin{equation}
\hat{\bsx} = \underset{\bsx} {\textrm{arg min}}~ - log \left(  P(\bsy|\bsx) \right) - log \left(  P(\bsx) \right),
\label{eq:logPostProb}
\end{equation}
where $f(\bsx) = - log \left(  P(\bsy|\bsx) \right)$ is the data fidelity term and $g(\bsx) = - log \left(  P(\bsx) \right) $ stands for the $log$-prior term of the image to estimate. 
With these notations, \eqref{eq:logPostProb} can be written as:
\vspace{-1mm}
\begin{equation}
\hat{\bsx} = \underset{\bsx} {\textrm{arg min}}~~ f(\bsx) + g(\bsx).
\label{eq:xhat}
\end{equation}
The noise Poisson probability density function is defined as
\begin{equation}
P(\bsy|\bsx) = \displaystyle\prod_i \frac{e^{-(\bsH \bsx)[i]}  {(\bsH \bsx)[i]}^{\bsy[i]}  }{\bsy[i]!},
\end{equation}
where $(\cdot)[i]$ represents the $i$-th component of a vectorized image. Thus the data fidelity term $f(\bsx)$ reads as
\begin{equation}
f(\bsx) = - \bsy^T log(\bsH \bsx) + {\boldsymbol{1}}^T \bsH \bsx + \text{constant},
\label{eq:datafidelity}
\end{equation}
where ${\boldsymbol{1}}$ is vector of length $n^2$ with each element equal to 1.

\vspace{-2mm}
\subsection{Quantum adaptative transform}
\label{sec:adaptrans}

A few attempts of applying tools from quantum physics to image processing can be found in the literature \cite{eldar2002quantum, gabbouj2013, youssry2015quantum, yuan2013quantum, LalegKirati15}. We focus hereafter on a specific application developed in \cite{dutta2020quantum, smith2018adaptive}, which uses quantum physics as a tool to obtain an adaptative basis which can be efficiently used to denoise an image. More specifically, we use the time-independent Schroedinger equation
\begin{equation}
 - \frac{\hbar ^2}{2m} \nabla ^2 \bpsi = - V(a)  \bpsi + E \bpsi,
\label{eq:schrodinger}
\end{equation}
which can be rewritten as an eigenvalue problem:
\begin{equation}
\bsH_{\mathcal{QAB}} \bpsi =  E \bpsi,
\label{eq:hamiltonian}
\end{equation}
where $\bsH_{\mathcal{QAB}} = - \frac{\hbar ^2}{2m} \nabla ^2 +V$ is the Hamiltonian operator, whose resolution gives a set of stationary solutions $\bpsi_i$ associated with eigenvalues (energies) $E_i$, normalized by $\int |\bpsi(a)|^2 da =1$. In physics, $V(a)$ is the potential where the quantum particle moves. Herein, we define it as the pixels' value. $\hbar$ representing the Planck constant and $m$ the mass of the particle in physics, are regrouped herein in a hyperparameter $\hbar^2/2m$. The solutions of \eqref{eq:hamiltonian} give oscillatory functions similar to the Fourier basis, with two main properties: i) the oscillation frequency increases with energy and ii) for the same basis function, the oscillation frequency depends on the pixels' value. The precise way the oscillation frequency depends on the pixels' values is regulated by the parameter $\hbar^2/2m$.
It was shown in \cite{dutta2020quantum, smith2018adaptive} that thresholding the image in this adaptative basis gives an efficient way to denoise it, in particular in the presence of \ab{Gaussian, Poisson or speckle} noise.


For imaging applications, the equation \eqref{eq:hamiltonian} has to be discretized. The corresponding Hamiltonian operator reads as:
\begin{eqnarray}
\bsH_{\mathcal{QAB}}[i,j]= \left \{
   \begin{array}{c c l}
      \bsx[i]+ 4 \frac{\hbar ^2}{2m} &  & for \; i=j,\\
       -\frac{\hbar ^2}{2m} & & for \; i = j \pm 1,\\
        -\frac{\hbar ^2}{2m} & & for \; i = j \pm n,\\
      0 & & otherwise,
   \end{array}
   \right.
\label{eq:H}
\end{eqnarray}
where $\bsx \in {\rm I \!R}^{n^2}$ is an image (\textit{i.e.,} $V=\bsx$) and $\bsH_{\mathcal{QAB}} [i,j]$ represents the $(i,j)$-th component of the operator $\bsH_{\mathcal{QAB}} \in {\rm I \!R}^{ n^2 \times n^2 }$. Note that zero-padding is used to handle the boundary conditions \cite{dutta2020quantum}. 
The set of $n^2$ eigenvectors corresponding to the Hamiltonian operator \eqref{eq:H}, which are primarily stationary wave solutions of \eqref{eq:schrodinger}, represents the adaptative transform and is denoted as the quantum adaptative basis (QAB) denoiser, $\mathcal{D_{QAB}}(\cdot)$, in the proposed algorithm.

In this context, the denoised image $\hat{\bsx}$ is retrieved by computing the projection coefficients $\balpha_i$ of the noisy image $\bsx$ onto the QAB, followed by a soft-thresholding $\tau_i$ and finally reverse projecting the refined coefficients, defined as:
\begin{eqnarray}
\hat{\bsx}= \sum _{i = 1} ^{n^2} \tau_{i}\balpha_{i}\bpsi_{i},
\label{eq:recons}
\end{eqnarray}
with,
$\tau_{i}= \left \{
   \begin{array}{c c l}
      1 &  & for \; i \leq s ,\\
      1 - \frac{i-s}{\rho} & & for \; i > s \;  and \; for \; 1 - \frac{i - s}{\rho} > 0 ,\\
      0 & & otherwise.
   \end{array}
   \right .$
where $s$ and $\rho$ are two thresholding hyperparameters.

Fundamentally, as stated above, these eigenvectors $\bpsi_i$ (called wave vectors in quantum physics) are oscillatory functions with an oscillation frequency typically proportional to the local value of $\sqrt{2m(E-V)}/\hbar$. Consequently, the same basis vector probes low potential regions with higher frequencies compared to a higher potential region. This adaptative nature of the basis vectors of $\mathcal{D_{QAB}}$ makes it very different from the Fourier and wavelet bases. The precise dependence of the local frequency on the pixels' intensity can be tuned through the value of the hyperparameter $\hbar ^2/2m$. As discussed in \cite{dutta2020quantum, smith2018adaptive}, the adaptative basis obtained from the raw noisy image is localized due to a subtle effect of quantum interference \cite{anderson1958absence}. In order to obtain a basis with extended vectors, which has been shown to be more efficient for denoising, one should compute the basis from an image obtained by first performing low-pass linear filtering with a Gaussian kernel of standard deviation $\sigma_{\mathcal{QAB}}$. This filtering is performed only to compute the most relevant adaptative basis, which is then used to denoise the original noisy image. 


\vspace{-2mm}
\subsection{Design of the QAB-PnP algorithm}
\label{sec:algo}
\vspace{-1mm}

The deconvolution problem \eqref{eq:xhat} can be addressed following the ADMM scheme \eqref{eq:step1}-\eqref{eq:step3} with the parameterization: $\boldsymbol{A} = - \boldsymbol{B} = \boldsymbol{I}_{n^2 \times n^2}$, $\boldsymbol{c} = \boldsymbol{0}_{n^2}$ ($\boldsymbol{I}_{n^2 \times n^2}$ is the identity matrix and $\boldsymbol{0}_{n^2}$ is a zero vector), giving at iteration $k$:
\begin{align}
&\bsx^{k+1}  = \underset{\bsx} {\textrm{arg min}}\Big(- \bsy^T log(\bsH \bsx) + {\boldsymbol{1}}^T \bsH \bsx \nonumber \\
& ~~~~~~~~~~~~~~~~~~~~~~~~~ + (\lambda^k/2) \norm{ \bsx - \bsz^k + \bsu^k }_2^2\Big)
\label{eq:admmPnP1}\\
& \bsz^{k+1}= \underset{\bsz} {\textrm{arg min}} \Big(g(\bsz) + (\lambda^k/2)\norm{ \bsx^{k+1} + \bsu^k - \bsz }_2^2 \Big)
\label{eq:admmPnP2}\\
& \bsu^{k+1} = \bsu^k + \bsx^{k+1} - \bsz^{k+1}.
\label{eq:admmPnP3}
\end{align}
To accelerate the convergence, the penalty parameter $\lambda$ is multiplied at each step by a factor $\gamma > 1$ \cite{chan2016plug}, instead of using a fixed value, \textit{i.e.,} $\lambda^{k+1} = \gamma \lambda^{k}$. One may observe that \eqref{eq:admmPnP2} can be associated with a denoising process. \ab{Several denoisers have been used in the literature, such as Gaussian denoisers combined or not with VST-like transforms. The main contribution of this work is to use $\mathcal{D_{QAB}}$ instead of classical denoisers.} Therefore while using the denoiser $\mathcal{D_{QAB}}$ as PnP denoiser, the step \eqref{eq:admmPnP2} becomes:
\vspace{-1mm}
\begin{equation}
\bsz^{k+1} =  \mathcal{D_{QAB}} \Big(\bsx^{k+1} + \bsu^k \Big).
\label{eq:admmPnP4}
\end{equation}

The convex problem \eqref{eq:admmPnP1} does not have an analytic solution. However, the gradient descent method \cite{ruder2016overview} offers an iterative way of solving it by calculating the gradient of the augmented Lagrangian $ \mathcal{L}_\lambda $, given by
\begin{equation}
\nabla_\bsx \mathcal{L}_\lambda = - \bsH^T \big(\bsy/(\bsH \bsx)\big) + \bsH^T {\boldsymbol{1}} + \lambda^k ( \bsx - \bsz^k + \bsu^k ),
\label{eq:admmPnP1Grad}
\end{equation}
where $ \nabla_\bsx $ represents the derivative with respect to $\bsx$ and $\bsy/(\bsH \bsx)$ stands for element-wise division.

\vspace{-3mm}

\begin{algorithm}[h!]
\begin{footnotesize}
\label{Algo:OMP}

\KwIn{ $\bsv$ , $\mathcal{T}$ , $\mathcal{D_{QAB}}$}

{\textbf{Initialization:}  $\bsr^0 = \bsv$ , $\Lambda^0 =  \emptyset$ , $\Phi^0$ is an empty matrix}\\


 \For{ $l$ from $0$ to $\mathcal{T} - 1$}{
 
 {$l = l + 1 $}\\
 
 {$\lambda^l = \underset{j = 1,2,...,\mathcal{T}} {\textrm{arg max}}~~ | \langle \bsr^{l-1},\bpsi_j \rangle |$, for $ \bpsi_j \in \mathcal{D_{QAB}}$ ~~~~~~~~~~(Break ties deterministically)}\\
 
 {$\Lambda^l = \Lambda^{l-1} \bigcup {\lambda^l}$ }\\
 
 {$\Phi^l = [\Phi^{l-1} ~~~~ \bpsi_{\lambda^l}]$}\\
 
 {$ a^l = \underset{a} {\textrm{arg min}}~~ \norm{ \bsv - \Phi^l a }_2$}\\

 {$ \bsr^l = \bsv - \Phi^l a^l$}
 }
\KwOut{$\hat{\balpha}$, which has nonzero elements only at $\Lambda^l$, \textit{i.e.,} $\hat{\balpha}_{\Lambda^l} = a^l$}
\caption{Modified OMP algorithm.}
\DecMargin{1em}
\end{footnotesize}
\end{algorithm}

\vspace{-9mm}

\begin{algorithm}[h!]
\begin{footnotesize}
\label{Algo:QAB}

\KwIn{ $\bsz$ , $\mathcal{D_{QAB}}$, $\mathcal{T}$, $s$ , $\rho$}

 {Compute the sparse coefficients $\hat{\balpha}_i$ with sparsity $\mathcal{T}$ by using the measurement data $\bsz$ and the operator $\mathcal{D_{QAB}}$ following Modified OMP method as illustrated in the Algorithm \ref{Algo:OMP}}\\
 
 {Threshold the coefficients  $\hat{\balpha}_{i}$}\\
 
 {Compute $\hat{\bsz}$ following \eqref{eq:recons}}
 
\KwOut{$\hat{\bsz}$}

\caption{QAB denoising algorithm.}

\DecMargin{1em}
\end{footnotesize}
\end{algorithm}

\vspace{-9mm}

\begin{algorithm}[h!]
\begin{footnotesize}
\label{Algo:QABPnPADMM}

\KwIn{ $\bsy$ , $\mathcal{E}$ , $\lambda_0$ , $\gamma$ , $\sigma_{\mathcal{QAB}}$ , $\frac{\hbar ^2}{2m}$ , $N$}

{\textbf{Initialization:}  $\bsx^0$ , $\bsz^0$ , $\bsu^0$}\\

{Compute a smooth version of $\bsy$ passing through a Gaussian filter with standard deviation $\sigma_{\mathcal{QAB}}$}\\
 
{Form the Hamiltonian matrix $\bsH_{\mathcal{QAB}}$ based on the smoothed version of $\bsy$ using \eqref{eq:H}}\\
 
 {Calculate the eigenvalues and eigenvectors of $\bsH_{\mathcal{QAB}}$}\\
 
 {Construct $\mathcal{D_{QAB}}$ using the eigenvectors $\bpsi_{i}$ of $\bsH_{\mathcal{QAB}}$}\\
 
 {Find total number of eigenvalue $\mathcal{T}$, less than the energy level $\mathcal{E}$ }\\

 \Begin{\textbf{ADMM process:}

 \For{ $k$ from $0$ to $N-1$}{
 
 {$ \bsx^{k+1} = \underset{\bsx} {\textrm{arg min}}~~ - \bsy^T log(\bsH \bsx) + {\boldsymbol{1}}^T \bsH \bsx + (\lambda^k/2) \norm{ \bsx - \bsz^k + \bsu^k }_2^2 $ }\\
 
 {$ \bsz^{k+1} = \mathcal{D_{QAB}}( \bsx^{k+1} + \bsu^{k} )$, following Algorithm \ref{Algo:QAB}}\\
 
 {$ \bsu^{k+1} = \bsu^{k} + \bsx^{k+1} - \bsz^{k+1}$}\\

 {$ \lambda^{k+1} = \gamma \lambda^{k}$}
 }

\KwOut{$\hat{\bsx} = \bsx^N$}
}

\caption{Proposed QAB-PnP algorithm.}

\DecMargin{1em}
\end{footnotesize}
\end{algorithm}
\vspace{-3mm}


In the proposed QAB-PnP algorithm, the denoising performed by $\mathcal{D_{QAB}}$ usually needs to execute the time-consuming task of calculating all the coefficients $\balpha_i$, despite the fact that very few coefficients actually participate in the restoration process, due to the soft-thresholding procedure. To decrease the computational load of the algorithm, it is convenient to focus only on basis vectors which contribute most (their total number will be denoted by $\mathcal{T}$) to $\mathcal{D_{QAB}}$. Since higher energy levels are above the threshold, it is sufficient to consider wave vectors up to an energy level $\mathcal{E}$, where $\mathcal{E}$ acts as a free hyperparameter.

The orthogonal matching pursuit (OMP) algorithm \cite{tropp2007signal} gives an efficient way of computing the most significant coefficients. It primarily aims at generating a sparse approximation $\hat{\balpha}_i$ with sparsity level $\mathcal{T}$ of the corresponding coefficients $\balpha_i$. Thus, OMP principally provides $\mathcal{T}$ non-zero coefficients $\hat{\balpha}_i$ such that an image $\bsv$ is approximated as $\bsv \simeq \mathcal{D_{QAB}} \hat{\balpha}_i$. 
To do so, firstly the most correlated column vector $ \psi_i \in \mathcal{D_{QAB}}$ is identified, followed by subtraction of its contribution and the process is restarted after the subtraction to obtain the second most important basis vector. The desired set of wave vectors is achieved after $\mathcal{T}$ iterations.
Note that in $\mathcal{D_{QAB}}$ the basis vectors are organized in the ascending order, so the first $\mathcal{T}$ basis vectors are mostly correlated with $\bsv$. Therefore, one can restrict OMP to the subset formed by the first $\mathcal{T}$ vectors, as illustrated in Algorithm \ref{Algo:OMP}.
The estimated sparse coefficients $\hat{\balpha}_i$ obtained from Algorithm \ref{Algo:OMP} are then used in the proposed QAB-PnP algorithm for image restoration. The Algorithms \ref{Algo:QAB} and \ref{Algo:QABPnPADMM} summarize the whole process.

\vspace{-1mm}
\section{Numerical experiments and results}
\label{sec:expe}

A detailed survey about the performance of the proposed QAB-PnP algorithm for Poisson image deconvolution is presented in this section through three simulations: one synthetic image and two cropped versions of standard images. All the images are distorted with a Gaussian blurring kernel  $h_{\sigma}^{4\times 4}$ of size $4 \times 4$ and standard deviation $\sigma = 3$. The study was conducted with three different Poisson noise levels corresponding to SNRs of 20, 15 and 10 dB. \ab{Note that the noise was image-dependent Poisson distributed and that the SNRs of the observations was computed a posteriori to emphasize the amount of noise}.

\vspace{-3.5mm}

\begin{table}[h!]
\begin{footnotesize}

\begin{center}
\caption{PSNR(DB)/SSIM (average over 200 noise realizations)}
\label{tab:tab_pip}
\vspace{-3mm}
\begin{tabular}{c c c c c}
\hline

\multirow{2}{*}{Sample} & \multirow{2}{*}{Method}
			 & \multicolumn{3}{c}{Poisson Noise}\\
			\cline{3-5}
			& & {\scriptsize SNR = 20dB} & {\scriptsize SNR = 15dB } & {\scriptsize SNR = 10dB }\\
						
\hline\hline

\multirow{6}{*}{Synthetic}
			
	& \multirow{2}{*}{TV-ADMM}
			& 26.46$\pm$0.10 & 24.80$\pm$0.34 & 22.52$\pm$1.55\\
	&		& 0.66$\pm$0.01 & 0.58$\pm$0.01 & 0.52$\pm$0.02\\
	\cline{2-5}
	& \multirow{2}{*}{P$^4$IP}
			& 23.90$\pm$1.37 & 20.91$\pm$2.18 & 18.96$\pm$3.34\\
	&		& 0.74$\pm$0.06 & 0.59$\pm$0.11 & 0.48$\pm$0.18\\
	\cline{2-5}
	& \multirow{2}{*}{QAB-PnP}
			& \textbf{29.86$\pm$0.12} & \textbf{27.18}$\pm$0.43 & \textbf{24.23}$\pm$1.34 \\
	&		& \textbf{0.92}$\pm$0.00 & \textbf{0.86}$\pm$0.01 & \textbf{0.74}$\pm$0.03\\
			
\hline

\multirow{6}{*}{Lena}

	& \multirow{2}{*}{TV-ADMM}
			& 27.37$\pm$0.31 & 24.52$\pm$0.65 & 19.97$\pm$1.32 \\
	&		& 0.74$\pm$0.01 & 0.66$\pm$0.01 & 0.52$\pm$0.02\\
	\cline{2-5}		
	& \multirow{2}{*}{P$^4$IP}
			& 27.32$\pm$0.44 & 24.87$\pm$2.76 & 18.67$\pm$4.83\\
	&		& \textbf{0.81}$\pm$0.01 & \textbf{0.76}$\pm$0.07 & 0.55$\pm$0.16\\
	\cline{2-5}
	& \multirow{2}{*}{QAB-PnP}
			& \textbf{28.97}$\pm$0.19 & \textbf{27.04}$\pm$0.44 & \textbf{20.18$\pm$3.39}\\
	&		& \textbf{0.81}$\pm$0.00 & 0.75$\pm$0.01 & \textbf{0.65}$\pm$0.08\\
\hline

\multirow{6}{*}{Fruits}
		
	& \multirow{2}{*}{TV-ADMM}
			& 20.51$\pm$0.38 & 19.02$\pm$0.23 & \textbf{17.54}$\pm$0.93 \\
	&		& 0.57$\pm$0.01 & 0.55$\pm$0.01 & 0.51$\pm$0.01\\
	\cline{2-5}	
	& \multirow{2}{*}{P$^4$IP}
			& 20.42$\pm$1.79 & 17.22$\pm$4.62 & 14.35$\pm$3.85 \\
	&		& 0.59$\pm$0.04 & 0.52$\pm$0.11 & \textbf{0.53}$\pm$0.04\\
	\cline{2-5}	
	& \multirow{2}{*}{QAB-PnP}
			& \textbf{21.37}$\pm$0.94 & \textbf{19.35}$\pm$0.96 & 17.28$\pm$3.55 \\
	&		& \textbf{0.62}$\pm$0.01 & \textbf{0.57}$\pm$0.02 & 0.51$\pm$0.12\\

\hline

\end{tabular}
\end{center}
\end{footnotesize}
\end{table}

\vspace{-4mm}

Poisson deconvolution is a widely studied field in the literature where PnP algorithms \ab{embedding a Gaussian denoiser combined or not with a VST} (\textit{e.g.}, BM3D) have shown promising performance \cite{rond2016poisson}.
The adaptative nature of our proposed scheme,  \ab{\textit{i.e.}, of the quantum denoiser to different noise statistics,} makes it well-adapted for the problem addressed and \ab{does not require using any additional transformation in the denoising step}.

\begin{figure}[h!]
\centering
\subfigure[\scriptsize{Lena image corrupted with 10 dB Poisson noise. The proposed algorithm is simulated with the hyperparameters $\mathcal{E} = 3.9$, $\lambda_0 = 1.5$, $\hbar ^2/2m = 4$.}]{\includegraphics[width=0.47\textwidth]{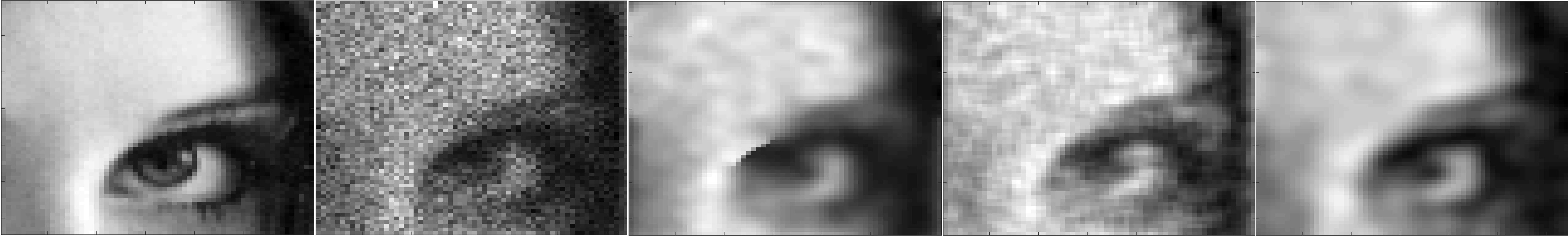}}\vspace{-3mm}

\subfigure[\scriptsize{Synthetic image corrupted with 15 dB Poisson noise. The proposed algorithm is simulated with the hyperparameters $\mathcal{E} = 4.1$, $\lambda_0 = 1.3$, $\hbar ^2/2m = 4$.}]{\includegraphics[width=0.47\textwidth]{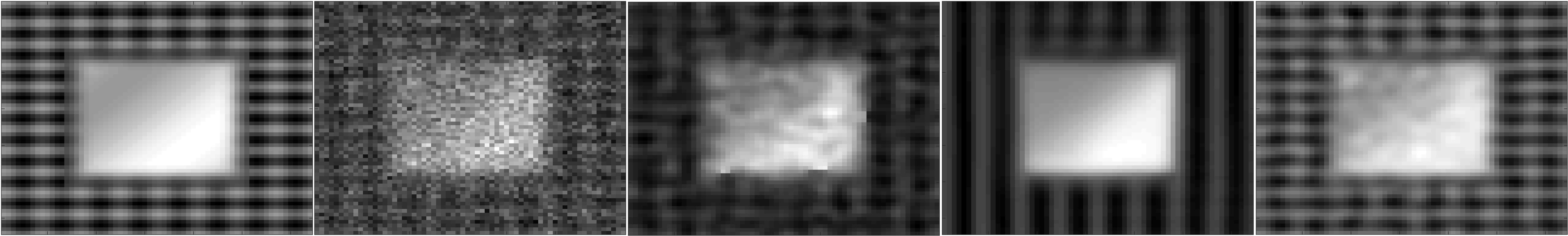}}\vspace{-3mm}

\subfigure[\scriptsize{Fruits image corrupted with 20 dB Poisson noise. The proposed algorithm is simulated with the hyperparameters $\mathcal{E} = 4.5$, $\lambda_0 = 3.15$, $\hbar ^2/2m = 4.3$.}]{\includegraphics[width=0.47\textwidth]{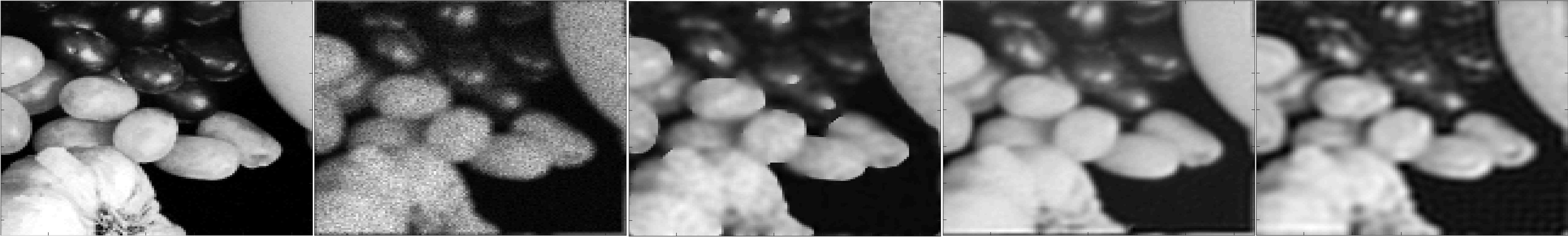}}
\vspace{-3mm}

\caption{In each row, the first, second, third, fourth and fifth images are accordingly clean, blurred noisy, deblurred result by TV-ADMM \cite{de2011alternating}, deblurred result by P$^4$IP \cite{rond2016poisson} and deblurred result by the proposed Algorithm \ref{Algo:QABPnPADMM}.} 
\vspace{-3mm}
\label{fig:results}
\end{figure}

To illustrate the practical interest of the proposed algorithm, we provide a comparison with a state-of-the-art PnP-ADMM methods called P$^4$IP in \cite{rond2016poisson} and a standard total-variation-based ADMM deconvolution algorithm adapted to Poisson observations in \cite{de2011alternating}, denoted hereafter by TV-ADMM. To ensure fair quantitative assessment of the restored images, the hyperparameters were tuned manually for all methods to obtain the best peak signal to noise ratio (PSNR) and structure similarity (SSIM) \cite{wang2004image}. Table \ref{tab:tab_pip} summarizes the average and standard deviation values for each set of experiments, obtained for all methods with 200 different noise realizations. For each set the best result is highlighted in bold. The quantitative results confirm that the proposed PnP scheme not only gives a better average value but also a smaller standard deviation compared to P$^4$IP, which highlights its good adaptability with high as well as low SNR images. For qualitative analysis, blurred Lena image observed through a Poisson process with SNR equal to 10 dB, a blurred synthetic image observed through a Poisson process with SNR 15 dB, and blurred fruits image observed through a Poisson process with SNR 20 dB are shown in Fig. \ref{fig:results}.

\begin{figure}[h!]
\centering
\includegraphics[width=0.2\textwidth]{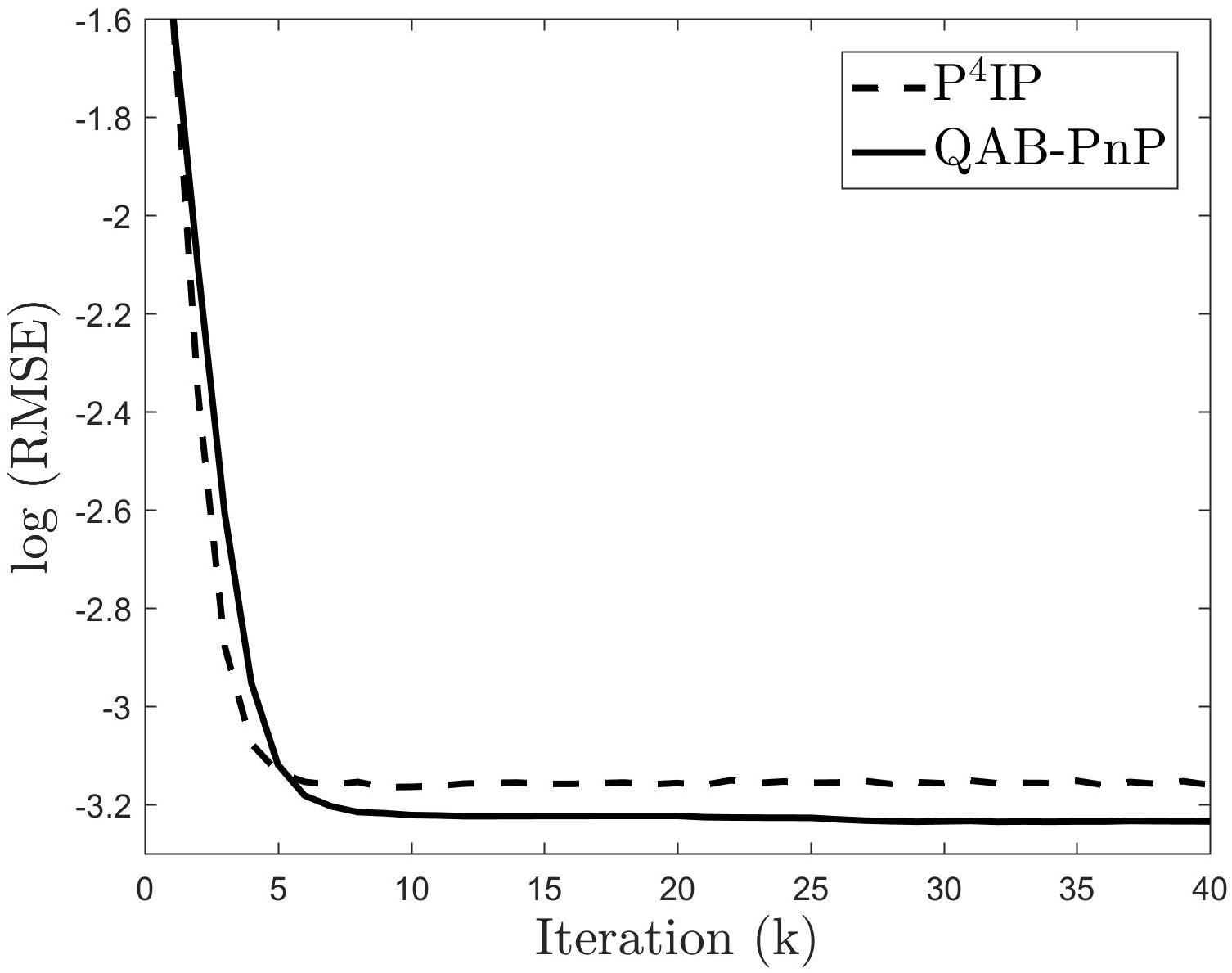}
\vspace{-3mm}
\caption{Logarithmic RMSE as a function of the number of iterations, performed on Lena image blurred by a Gaussian blurring kernel $h_{\sigma = 3}^{4\times 4}$ and contaminated by Poisson noise with a SNR of 20 dB.}
\vspace{-6mm}
\label{fig:p4ipvsprop}
\end{figure}

To confirm the interest of considering OMP with hyperparameter $\mathcal{E}$,  Table \ref{tab:com_time} illustrates the average processing time and the PSNR of the proposed algorithm with and without OMP, \textit{i.e.} with and without using the hyperparameter $\mathcal{E}$ for the synthetic image. This reveals that the gain in computational time is significant with very small accuracy loss. Finally, Fig. \ref{fig:p4ipvsprop} shows the behavior of the proposed algorithm compared to P$^4$IP.


\vspace{-3.5mm}
\begin{table}[h!]
\begin{footnotesize} \begin{center}
\caption{Average computational time and PSNR (for synthetic image)}
\vspace{-2mm}
\label{tab:com_time}
\begin{tabular}{c || c  c}
Simulation & With OMP, best $\mathcal{E}$ & Without OMP \\
\hline
Run time (sec)		& 40.575 & 214.394 \\

PSNR	 (dB)			& 29.86  & 30.08 \\
\end{tabular} \end{center} \end{footnotesize}
\end{table}

\vspace{-4mm}

\section{Conclusions}
\label{sec:conclusion}

A new adaptable PnP scheme inspired by quantum mechanical tools has been studied in this work to handle Poisson inverse problems. 
The proposed algorithm uses a quantum mechanics-based QAB denoiser, whose design makes it well-adapted to \ab{different noise statistics}, \ab{explaining its good behavious as denoiser embedded in a PnP-ADMM algorithm}. The numerical experiments reveal the potential of the proposed algorithm in the Poisson deconvolution context, even in the low SNR case, where VST-based models have very irregular performances, as highlighted by the high standard deviation values for P$^4$IP algorithm in Table \ref{tab:tab_pip}.

\bibliographystyle{IEEEbib}
\bibliography{ADMMPnPcon}

\end{document}